\documentclass[iop]{emulateapj}

\def\deg{$^{\circ}$}
\def\radm2{${\rm rad\,m}^{-2}$}

\begin{document}
\title{~~\\ ~~\\ Are There Rotation Measure Gradients Across AGN Jets?}
\shorttitle{Looking for Gradients}
\author{G. B. Taylor\altaffilmark{1}, and R. Zavala\altaffilmark{2}}
\email{gbtaylor@unm.edu}
\altaffiltext{1}{Department of Physics and Astronomy, University of New
Mexico, Albuquerque NM, 87131, USA; gbtaylor@unm.edu. Greg Taylor is also an
Adjunct Astronomer at the National Radio Astronomy Observatory.}
\altaffiltext{2}{U.S. Naval Observatory, Flagstaff Station, 10391 
W. Naval Observatory Rd. Flagstaff AZ, 86001, USA}

\slugcomment{As accepted to ApJ Letters}

\begin{abstract}

We report on multi-frequency polarimetry Very Long Baseline
Interferometry observations of active galactic nuclei using the VLBA.
These observations are used to construct images of the Faraday
Rotation Measure (RM) in J1613+342, Mrk\,501, 3C\,371, and BL Lac.
Despite having resolved the jets in total intensity and polarization
for three of these sources no RM gradients are found.  This is in
contrast to the large fraction of sources with RM gradients now
claimed in the literature, and invoked as evidence in support of
helical magnetic fields.  We propose objective criteria for 
establishing what constitutes an RM gradient.  Furthermore, although
we note the absence of simple, monotonic gradients, comparison
with simulations could reveal systematic changes in 
the RM which may be masked by a varying jet orientation.

\end{abstract}


\section{Introduction}

Understanding how the jets from active galactic nuclei (AGN) are
launched is an outstanding question in astrophysics. Blandford \&
Znajek (1977) proposed an electro-magnetic model by which the energy
of the black hole could launch a relativistic jet. Magnetic fields
play an important role in this and many of the proposed models and
simulations (e.g., Meier, Koide \& Uchida 2001, 
Meier 2005).  A predicted consequence of strong, ordered magnetic
fields that wrap around the jet is a gradient in the Faraday Rotation
Measure (RM) transverse to the long axis of the jet (Blandford 1993).
Asada et al. (2002), and Zavala \& Taylor (2005), found evidence for
RM gradients in 3C\,273, and this was quickly followed with claims for
RM gradients in 0745+241, 0820+225, Mrk\,501, and 3C\,371 by Gabuzda
et al. (2004), and in DA\,237 and 1156+295, 1749+096 by Gabuzda et
al. (2008). Recently, Contopoulos et al. (2009) claim that RM
gradients have been established in 36 instances in 29 sources and
further suggest that the sense of RM gradients has a preferred
direction. They claim that the preponderance of clockwise gradients
(22) over counterclockwise (14) gradients is the result of an
invariant twist of the magnetic fields in AGN jets. Note, however that
K{\"o}nigl (2010) shows that, if such a preference exists, it can be
explained in terms of standard models. 

Considering the observational results one is left with 
the impression that transverse RM gradients on parsec scales 
are the rule rather than the exception.  The best
established (spatially resolved) case is still that in 3C\,273, 
which is an exceptionally nearby quasar, and one of the brightest 
radio sources in the sky at centimeter wavelengths. With the 
exception of  3C\,273, many of the transverse RM gradients claimed 
in the literature are in sources where the synthesized beam spans not 
much more than the width of the jet (e.g., Gabuzda et al. 2004). Increasing 
the angular resolution by going to shorter wavelengths is seldom rewarding, 
as the jets are steep spectrum so at shorter wavelengths one 
tends to probe only the brightest part of the jet base. Observers thus 
run the risk of extrapolating results from a nearby, well-resolved 
object onto a less well-resolved population. A further 
difficulty is that the observations require a broad coverage in 
wavelength-squared space, obtained simultaneously, with matching resolution. This wavelength 
coverage is not always achieved in VLBI polarization observations. 

The considerable body of observational literature describing RM features 
on parsec scales begs for support from simulations. Given a relativistic 
jet pointed close to the line-of-sight with a dominantly toroidal magnetic 
field component that extends into a Faraday rotating plasma, what 
should observers see? 
Broderick \& McKinney (2010) have recently completed such an
effort. Their results show that jets that are not well resolved will
not produce simple transverse RM gradients. Broderick \& McKinney also
verify the observational intuition that spectral index effects will
significantly complicate the interpretation of RM gradients in
optically thick regions. They also provide a physical connection
between the black hole and the RM via the accretion rate.

We examine transverse RM gradients in relativistic jets which 
maximize the spatial resolution at our disposal by
selecting a sample of 4 strong, broad
jets from the MOJAVE sample (Lister et al. 2009) 
known to be well polarized at centimeter wavelengths.  We then
carried out sensitive multi-band VLBI polarimetry on them. 
We present these observations and note the 
presence or absence of transverse RM gradients. We conclude with a 
set of objective observational criteria for determining if a transverse 
RM gradient exists.

We assume $H_0$ = 71 km s$^{-1}$ Mpc$^{-1}$ and a $\Lambda$CDM cosmology
with $\Omega_\lambda=0.7$ and $\Omega_m=0.3$ (e.g. Eisenstein et al. 2005; Hinshaw et al. 2009).

\section{Observations and Data Reduction}

The observations were carried out at 8, 13, and 15 GHz on 2003
September 26 using all ten elements of the VLBA of the NRAO.  Each
source was observed in $\sim$12 scans of 2-3 minutes each at each
frequency band.  Amplitude calibration
for each antenna was derived from measurements of antenna gain and
system temperatures during each run. Delays between the stations'
clocks were determined using the AIPS task FRING (Schwab \& Cotton
1983). Calibration was applied by splitting the multi-source data set
immediately prior to preliminary editing, imaging (using natural weighting), 
deconvolution (using the CLEAN algorithm) and
self-calibration in {Difmap} (Shepherd, Pearson, \& Taylor 1995).
The preliminary models developed in
{Difmap} were subsequently applied in AIPS to make phase corrections,
to determine the leakage terms between the RCP and LCP feeds and to
correct for residual phase differences between polarizations.  Total 
intensity images were made using all available IFs near 
8, 13 or 15 GHz bands, while Stokes Q and U images were constructed
at 8.114, 8.209, 8.369, 8.594, 12.915, 13.885, 14.915, and 15.391 GHz
with 8 MHz bandwidths.
Further subdivision of the frequencies within the 8 MHz bandwidths
was deemed unnecessary
given the maximum RMs observed.
Absolute calibration of the polarization angle was accomplished by 
observations of 3C279, 1751+096, and BL Lac, which are observed regularly
by the VLA polarization monitoring 
program\footnote{http://www.aoc.nrao.edu/~smyers/calibration/} to 
provide calibration information for VLBA users.
Rotation measure (RM) maps were then formed from matched resolution 
polarization angle images at these 8 frequencies.  Resolutions were 
matched by applying the appropriate taper in the $(u,v)$ plane 
when imaging and then
restoring with a common beam size.   Pixels in the RM
images were blanked if the error in any polarization angle image 
exceeded 20 degrees, or if the total intensity was less than 6 times
the rms noise level. No correction for redshift has been made so the 
intrinsic rotation measures in the rest-frame of the source are larger
than the observed values by $(1+z)^2$.

\section{Results}

For each source we examined the RM image and took slices across the
jet. Outside of the core region, the RMs are well fit by a
$\lambda^2$-law fit, and have constant fractional polarization,
which taken together indicates that their origin is most likely in an
external Faraday screen, and not due to thermal material intermixed
with the radiating plasma (Burn 1966). In the core region some large
deviations from a $\lambda^2$-law may be present due to unresolved
substructures. For one source (3C\,371) RMs could only be determined at
the base of the jet where the jet is only marginally resolved. For
the other three sources we were able to resolve the jet and we present
the RM images obtained.


\subsection{J1613+3412}

This source is a flat spectrum radio quasar (FSRQ) with a 
redshift of z=1.401 (Burbidge 1970). 
It emits powerfully across the electromagnetic 
spectrum and is detected in the Gamma-rays by EGRET (Hartman et al. 1999). 
It is detected by $Fermi$ with a flux of 
0.5 $\times 10^{-9}$ (100 MeV to 100 GeV) photons cm$^{-2}$ sec$^{-1}$
(Abdo et al. 2010).  It is
a known superluminal source (Piner \& Kingham 1997; Lister et al. 2009)
with apparent velocities between 9 and 15 c. 
A compact, flat spectrum core is identified with a steep spectrum
jet extending to the south.  This jet is exceptionally
broad, with an opening angle of $\sim$60\deg. Piner \& Kingham (1997) estimate
the orientation of the jet at $<$7\deg\ to the line-of-sight.  

In Fig.~1 we show the RM image for J1613+3412 with an inset slice
across the jet at its broadest point. The jet is $\sim$6 mas wide or
about 5 beams across, so well resolved.  We find a Spearman correlation
coefficient of $-$0.2, and a reduced
$\chi^2$ for a linear fit of 2.8 which reject a linear
gradient across the jet. There is a sharp change in the RM along the jet
from relatively high values in the core of $-$600 \radm2\ decreasing
to an average of 0 \radm2\ in the outer part of the jet (Spearman 
correlation $-$0.83). This
behavior is typical for blazars which have much higher RMs in their
central regions, presumably because of the increase in density near
the bottom of the gravitational potential of the supermassive
black-hole (Zavala \& Taylor 2002, 2003, 2004). The RM image from Zavala \&
Taylor (2003) is quite similar though it doesn't cover as much of the
jet.

\begin{figure}
\plotone{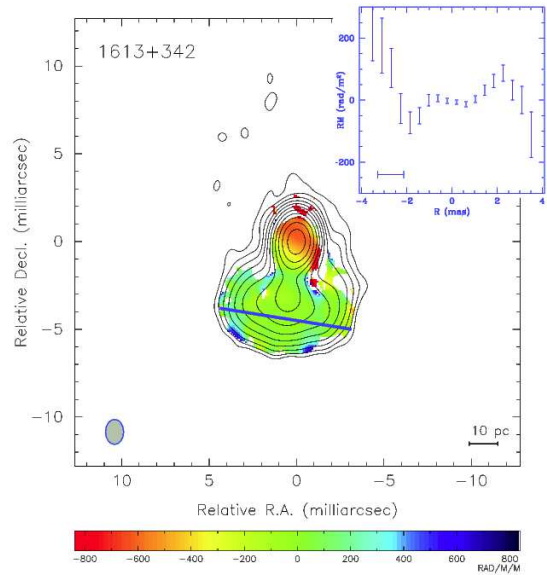}
\caption{Rotation Measure image of J1613+342 in false color with 
a slice perpendicular to the jet axis inset.  The zero point on the 
inset slice corresponds to the center of the jet, and the bar denotes
the angular resolution.  Image coordinates are only relative since 
no phase referencing was performed.
The synthesized beam (shown in lower left) has dimensions 
1.6 $\times$ 1.2 mas in position angle 0\deg.
Contours are drawn from the 8 GHz total intensity 
image logarithmically at factor 2 intervals with the first contour at 
2 mJy/beam. The Spearman correlation coefficient is $-$0.2, and the reduced
$\chi^2$ for a linear fit of the RM with position is 2.8.}
\end{figure}

\subsection{Mrk 501}

Mrk 501 is a well-studied BL Lac object at z=0.0337 (Ulrich et al. 1975). 
It also emits across the electromagnetic spectrum, and was one of the first
sources detected at TeV energies (Quinn et al. 1996). 
It is detected by $Fermi$ with a flux of 
8.3 $\times 10^{-9}$ (100 MeV to 100 GeV) photons cm$^{-2}$ sec$^{-1}$
(Abdo et al. 2010). VLBI observations have revealed a 
limb-brightened structure, possibly the result of a dual velocity structure 
(Giroletti et al. 2004, 2008). Giroletti et al. (2008) find a well-ordered 
magnetic field structure in the outer jet (100 mas from the core) 
suggesting that the RM in this region is low and uniform. 
The maximum jet speed in
the MOJAVE archive is 0.2 c (Lister et al. 2009), though Piner et al. (2009)
report a speed of 3.3 $\pm$ 0.3 c using higher resolution 43 GHz observations.

In Fig.~2 we show the RM image for Mrk 501 with an inset slice across
the jet at its broadest point where the RMs are well determined,
roughly 10 mas from the core. The jet is $\sim$8 mas wide or about 6
beams across, so well resolved, although the detection of polarized
flux is restricted to the brighter $\sim$75\% of the jet. The slice
across the jet shows some variations between 0 and 200 \radm2, but 
no gradient (Spearman correlation coefficient is $-$0.1, and the reduced
$\chi^2$ for a linear fit is 2.0). Gabuzda et al. (2004) claim a gradient 
from $-$63 $\pm$ 30 to $+$130 $\pm$ 20 \radm2 at a location somewhat
closer to the core than the slice shown in Fig.~2. The RM determination of Gabuzda
et al. was based on three observations of the polarization angle 
taken at 5, 8 and 15 GHz in 1997 using the standard frequency bands. 
Their resolution was somewhat lower (beam 2.55 $\times$ 1.89 mas in 
position angle $-$24\deg). Croke, O'Sullivan \& Gabuzda (2010) 
report a transverse RM gradient based on 1.6 to 8.5 GHz 
VLBA observations. According to their spectral index maps the
location where the RM gradient is detected is relatively flat.  

We see a clear change in RM along the jet from relatively high values in the
core of 2000 \radm2\ decreasing to an average of 10 \radm2.  In the
core region there are large and abrupt changes in the RM, however we
caution against interpreting this as an RM gradient since substructure
in the core is likely. Furthermore the local jet direction is not
always well defined and one can find
abrupt changes in RM in essentially every orientation in the cores of
AGN.

\begin{figure}
\plotone{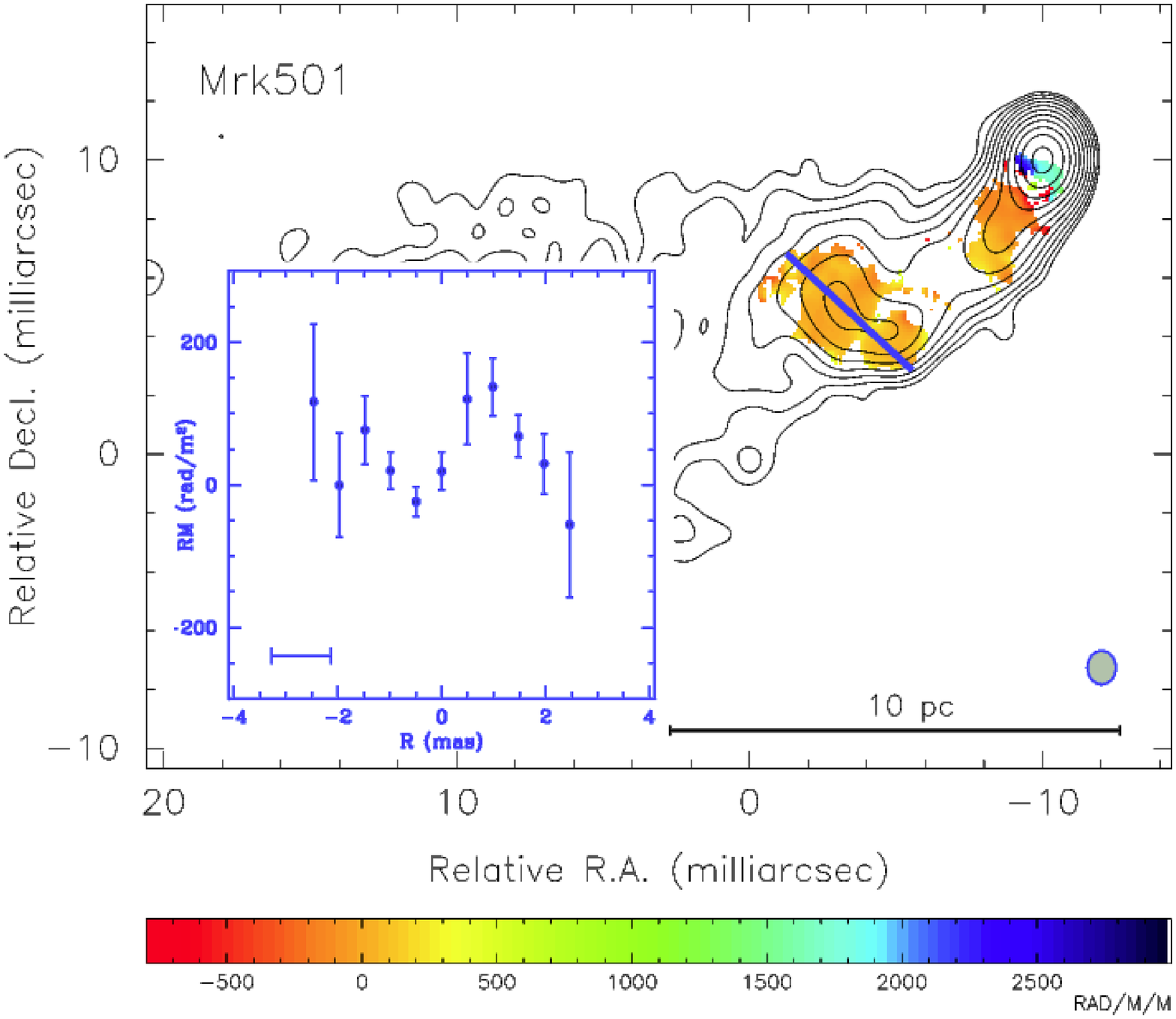}
\caption{Rotation Measure image of Mrk 501 in false color with 
a slice perpendicular to the jet axis inset. 
The synthesized beam has dimensions 
1.4 $\times$ 1.2 mas in position angle 0\deg.
Contours are drawn starting at 
1 mJy/beam. The Spearman correlation coefficient is $-$0.1, and the reduced
$\chi^2$ for a linear fit is 2.0.}
\end{figure}

\subsection{3C\,371}

3C\,371 is a high power BL Lac object at a redshift of 0.051 
(Lawrence et al. 1996). The radio jet has been 
well studied on the parsec scale by G{\'o}mez \& Marscher (2000)
and they  suggest a classification as a transition object between a 
BL Lac and a radio galaxy. 
3C\,371 is detected by $Fermi$ with a flux of 
1.9 $\times 10^{-9}$ (100 MeV to 100 GeV) photons cm$^{-2}$ sec$^{-1}$
(Abdo et al. 2010). The jet is relatively slow moving (0.1c, 
Lister et al. 2009) and taking into account the many MOJAVE 
epochs and two epochs of G{\'o}mez \& Marscher (2000) is 
consistently weakly polarized. The kiloparsec scale jet has been 
detected in the radio (Cassaro et al. 1999), optical (Scarpa et 
al. 1999) and in X-rays (Pesce et al. 2001). Taken together 
these results suggest an object with radio galaxy like properties 
(Zensus 1997) although variability undoubtedly plays a role.
Of the four sources in this work, 3C\,371 may have the jet with 
the largest angle to the line of sight of $\approx$ 20\deg\ 
(G{\'o}mez \& Marscher, 2000).

In Fig.~3 we show the RM image for 3C\,371 with an inset slice across
the region where the RMs cover the largest extent. The jet, however,
is not resolved so we can say little about RM gradients in this 
object. Further out the jet is well resolved in total intensity, 
but is not detected in linear polarization.

\begin{figure}
\plotone{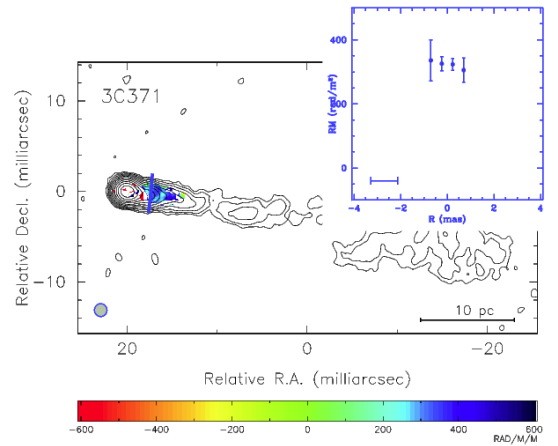}
\caption{Rotation Measure image of 3C\,371 in false color with 
a slice perpendicular to the jet axis inset. 
The synthesized beam has dimensions 
1.25 $\times$ 1.25 mas. Contours are drawn starting at 
1 mJy/beam. There is insufficient data to compute the Spearman correlation 
coefficient or the $\chi^2$ for a linear fit. }
\end{figure}

\subsection{BL Lac}

The archetype of its class (Strittmatter et al. 1972), BL Lac is at
z=0.0686 (Vermuelen et al. 1995) and well known for its optical variability and
powerful emission across the electromagnetic spectrum (Bregman et al. 1990),
including $\gamma$-rays (Bloom et al. 1997). The compact
structure of BL Lac has been studied extensively with VLBI polarimetry
by Denn, Mutel \& Marscher (2000) who found highly polarized components were
ejected from the core and moved away on helical trajectories. BL Lac
is also a known superluminal source with velocities of 3.5 c
(Mutel et al. 1990) to as high as 10 c with a jet closely aligned 
to the line of sight (Jorstad et al. 2005).

In Fig.~3 we show the RM image for BL Lac with an inset slice across
the jet at its broadest point where the RMs are well determined,
roughly 5 mas from the core. The jet is $\sim$4 mas wide or about 3
beams across, so resolved. The slice across the jet shows lower RMs
at the edges of the jet and an RM of $-$250 in the center. The error
in the RMs at the edge of the jet are large and there are clearly some
values along the edge at low SNR that should not be trusted. The RMs
in the core are low ($-$70 $\pm$ 10 \radm2) which is a bit unusual for
AGN, but we could just be catching BL Lac in a low state. Zavala \&
Taylor (2003) found an RM of $-380$ \radm2, and before that Reynolds,
Cawthorne \& Gabuzda (2001) found a core RM of $-$550 \radm2.

\begin{figure}
\plotone{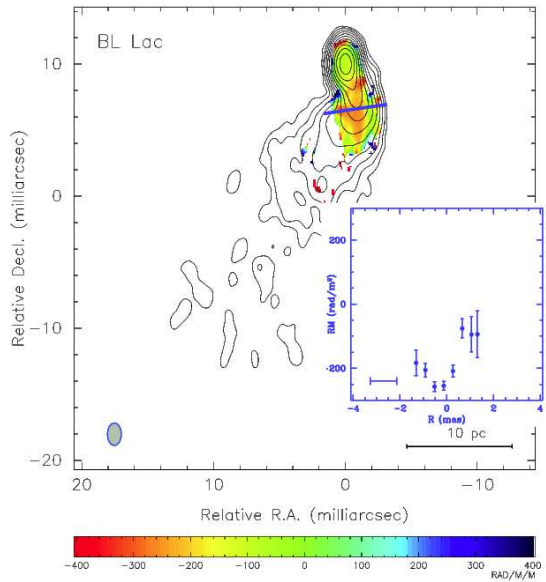}
\caption{Rotation Measure image of BL Lac in false color with 
a slice perpendicular to the jet axis inset. 
The synthesized beam has dimensions 
1.7 $\times$ 1.0 mas in position angle 0\deg. 
Contours are drawn starting at 
1 mJy/beam. The Spearman correlation coefficient is 0.5, and the reduced
$\chi^2$ for a linear fit is 5.7.}
\end{figure}

\section{Discussion}

Contopoulos et al. (2009) claim to have found 29 AGNs from the 
literature that have RM
gradients transverse to the jet direction that are monotonic and
``extend across all or nearly all of the jet''. Contopoulos et
al. don't present any statistics on how well fit the gradients are
by a straight line.  This list includes both Mrk 501 and 3C\,371
studied here.  The observations of Gabuzda et al. (2004) 
use three frequencies which is essentially the minimum with which 
to claim a rotation measure determination and the observations 
(with the possible exception of Mrk 501) do not significantly resolve the 
region in which the RM is detected. For example, approximately one
synthesized beamwidth spans the RM extent of 3C\,371 in Figures 1 and 2 
of Gabuzda et al. (2004). Contopolous et al. (2009) also claim the
presence of RM gradients in B0212+735, and B1803+784 based on the
observations of Zavala \& Taylor (2003). In both of these instances
the jets were not resolved and it appears that spurious values at edge
pixels (a common problem in determinations of RMs in regions of 
low signal-to-noise) were over-interpreted to indicate the presence of a gradient.

Contopoulos et al. list 7/29 sources with RM gradients
at a distance of 0 mas and thus coincident with the VLBI
core. As the simulations of Broderick \& McKinney (2010)
show this is a problematic area for RM measurements.
High spatial resolution polarimetry observations reveal the
complex nature of these regions and further justify
caution in claiming detection of RM gradients at the
``core'' of AGN (Zavala \& Taylor 2001; Denn, Mutel \& Marscher 2000).

If one only has two measurements then a gradient will exist anytime
these values differ. We suggest that to establish an RM gradient requires
that the following two observational and two physical constraints are met:
\begin{itemize}
\item[1.]{At least three resolution elements across the jet.}
\item[2.]{A change in the RM by at least three times the typical error.}
\item[3.]{An optically thin synchrotron spectrum at the 
location of the gradient.}
\item[4.]{A monotonically smooth (within the errors) change in the RM 
from side to side.}
\end{itemize}

To date the above criteria have been met (to our knowledge) only for
3C\,273. In this work we attempt to improve our observational footing
by applying these criteria to four additional sources. While 3C\,371
fails on the first (observational) test, the other three sources pass
the first two observational criteria, and the third physical criterion
of an optically thin jet, but fail on the critical fourth physical
criterion.  That is, these three sources have data of sufficient
quality that they could reveal an RM gradient, but no smooth monotonic
gradient across the jet is found.  However, even if simple monotonic
gradients across AGN jets turn out to be rare, it is still important
to make careful comparison of the observed RM profiles with
simulations.  For example, the inset slice in our Fig.~1 bears 
some resemblance to Fig.~5a of Broderick \& McKinney (2010).
In their simulation the jet is oriented at 10 degrees to the
line of sight which is very close to the estimate of 7 degrees
(Piner \& Kingham 1997). 

It is worth noting in passing that all of the sources studied
here are detected in $\gamma$-rays. This is also true for
3C\,273. This may be the result of the selection criteria of broad
jets, together with the fact that sources with broad jets tend to be
preferentially detected at high energies (Taylor et al. 2007, Linford
et al. 2010).

\section{Conclusions}

Based on well-resolved rotation measure imaging of three AGN we find
that none of them exhibit a monotonic linear gradient in the RM across
the jet.  Examination of claims for RM gradients in the literature
(including two of the four sources we observed) are found to be based
on observations that generally have not resolved the jet, include
components with large opacities, or suffer from spurious edge effects.
We offer a set of criteria that can be used to reliably determine if
an RM gradient exists.  Moreover, even if simple linear gradients
across AGN jets turn out to be rare, it will be important to make
careful comparison of the observed RM profiles with simulations (e.g.,
Broderick \& McKinney 2010).  Future observations with higher
resolution, such as will be afforded by VSOP-2, will be crucial in
order to increase the number of parsec-scale jets that are well
resolved across the jet.  New wideband correlators will allow for RM
synthesis (Brentjens \& de Bruyn 2005) observations with increased precision and the ability to
disentangle multiple RM screens.  Another possibility, assuming
improvements in the sensitivity of VLBI arrays, is the
detection of polarization in two-sided jet sources such as NGC 1052
(Vermuelen et al.  2003) or 3C\,84 (Taylor et al.\ 2006). Polarized
counter-jets may reveal RM gradients consistent with a dominant
toroidal field (Blandford 1993).

\acknowledgments

We thank the staff of the U.S. Naval Observatory Library and 
especially U. Grothkopf, librarian at the  European Southern 
Observatory, Garching, for assistance in our literature
search. The National Radio Astronomy Observatory is operated by
Associated Universities, Inc., under cooperative agreement with the
National Science Foundation. This research has made use of data from 
the MOJAVE database that is maintained by the MOJAVE team (Lister et al.\ 
2009), NASA's Astrophysics Data System Bibliographic Services and 
the NASA/IPAC Extragalactic Database (NED) which is operated by the Jet 
Propulsion Laboratory, California Institute of Technology, under contract 
with the National Aeronautics and Space Administration.

{\it Facilities:} \facility{VLBA ()}

\clearpage

\def\dg{$^{\circ}$}

\end{document}